\documentstyle[aps,prb,multicol,epsf]{revtex}

\tighten
\begin{document}
\draft

\title{Competition between ferromagnetic and charge-orbital ordered
  phases \\ in Pr$_{1-x}$Ca$_{x}$MnO$_3$ for $x$=1/4, 3/8, and 1/2}

\author{Takashi Hotta and Elbio Dagotto}

\address{National High Magnetic Field Laboratory, Florida State
  University, Tallahassee, Florida 32306} 

\date{\today}

\maketitle

\begin{abstract}
Spin, charge, and orbital structures in models for doped manganites
are studied by a combination of analytic mean-field and numerical
relaxation techniques. 
At realistic values for the electron-phonon and antiferromagnetic
$t_{2g}$ spin couplings, a competition between a ferromagnetic (FM)
phase and a charge-orbital ordered (COO) insulating state is found for
$x$=1/4, 3/8, and 1/2, as experimentally observed in
Pr$_{1-x}$Ca$_{x}$MnO$_3$ for $x$=0.3$\sim$0.5.
The theoretical predictions for the spin-charge-orbital ordering pattern 
are compared with experiments.
The FM-COO energy difference is surprisingly small for the densities
studied, result compatible with the presence of a robust 
colossal-magnetoresistive effect in Pr$_{1-x}$Ca$_{x}$MnO$_3$ in a
large density interval.
\end{abstract}


\begin{multicols}{2}
\narrowtext

The explanation of the colossal magnetoresistance (CMR) phenomenon in
perovskite manganese oxides is currently one of the most challenging
problems in condensed matter physics.\cite{review}
In addition to their unusual magneto-transport properties, manganites
present a very rich phase diagram involving phases with spin, orbital,
and charge order. 
Early investigations of these materials relied on the
``double-exchange'' (DE) mechanism for ferromagnetism,
in which $e_g$ electrons optimize their kinetic energy
if the $t_{2g}$ spins background is polarized.
However, this simple picture is not sufficient to rationalize the CMR 
effect since the properties of the insulating state involved in the
metal-insulator transition play a key role.
In fact, the understanding of the A-type antiferromagnetic (AFM)
insulator LaMnO$_3$ requires a two orbital model and strong
electron-phonon or Coulomb interactions to induce the complex
spin and orbital arrangement characteristic of this state.\cite{Hotta1}
The evolution of the undoped phase with light hole doping, and its
eventual transformation into a ferromagnetic (FM) metal,\cite{LSMO} is
nontrivial and it is expected to proceed through a mixed phase process
involving nanometer domains.\cite{moreo}
The conspicuous percolative characteristics of the transition have
been recently explained by the influence of quenched disorder in the
hopping and exchanges of manganite models.\cite{random} 

At high hole densities near
$x$=1/2 a strong CMR phenomenon occurs.\cite{Tokura} 
For example, in  Nd$_{1/2}$Sr$_{1/2}$MnO$_3$,\cite{NSMO} a low
magnetic field of a few Teslas is enough to induce a metal insulator
transition between the FM metallic and charge-ordered (CO) insulating
phases.
Based on recent computational studies at $x$=1/2, that clearly showed 
the presence of a first-order level-crossing transition between the FM
and CE-type CO states,\cite{recent} 
a simple picture to explain this result can be constructed. 
At densities slightly above the critical hole doping $x_{\rm c}$ $\sim$1/2 of
the FM-CO transition it is reasonable to expect that small magnetic
fields will transform the ground state from CO to FM, since these
states are close in energy.
However, the FM metallic and CO insulating phases have a quadratic and
linear dependence with $x$, respectively, and  small magnetic fields
will become rapidly ineffective to produce such a transition as $x$
grows away from $x_{\rm c}$. Thus, in this simple scenario the CMR
effect can only occur in a narrow density window around $x_{\rm c}$.
Unfortunately, some experiments suggest that this scenario is 
incomplete. In fact, it is well-known that for Pr$_{1-x}$Ca$_{x}$MnO$_3$
(PCMO)\cite{PCMO} the CMR effect occurs in a $wide$ density range
0.3$\alt$$x$$\alt$0.5, and 
at first sight it appears that a simple level-crossing picture is not
suitable for this compound.

It is the purpose of this paper to discuss a possible alternative
explanation for the CMR phenomenon in PCMO. 
In the new scenario it is still claimed that at a fixed density
small magnetic fields can induce FM-CO transitions as in 
the simple level crossing picture, but the key new idea is that
different types of CO phases are stabilized at different densities.
The calculations are here  carried out at some 
special dopings, $x$=1/4, 3/8, and 1/2, for technical reasons to be
discussed below, and using the two-orbital model strongly coupled to
the Jahn-Teller (JT) distortions. 
A remarkable result observed in the present study is that, for
realistic parameters, the region of competition between FM and CO
phases is found to occur virtually independently of $x$, namely for
the same values of couplings. 
The FM-CO energy difference is surprisingly small, as long as the CO
state is optimized at each density. 
These novel candidate CO phases competing with the FM state are
here described in detail. 
In previous literature it is usual to find references to these states
as ``$x$=1/2 CO plus defects'', due to their experimentally observed 
similarities with the
$x$=1/2 charge arrangement, although they have different electronic
densities. Here concrete examples of CO states with $x$$\neq$1/2 are
presented. 
The discussion below lead us to believe that similar
conclusions would 
be obtained if Coulomb interactions were included in the calculations.

Let us consider the hopping of $e_{\rm g}$ electrons, tightly coupled to
localized $t_{\rm 2g}$ spins and the JT distortions of the MnO$_6$
octahedra. Their Hamiltonian is 
\begin{eqnarray}
 H \! &=& \! \!-\! \sum_{{\bf ia}\gamma \gamma' \sigma}
 t^{\bf a}_{\gamma \gamma'} c_{{\bf i} \gamma \sigma}^{\dag} 
 c_{{\bf i+a} \gamma' \sigma}  \!-\! J_{\rm H} \sum_{\bf i}
 {\bf s}_{\bf i} \cdot {\bf S}_{\bf i} 
 \!+\! J' \sum_{\langle {\bf i,j} \rangle}
 {\bf S}_{\bf i} \cdot {\bf S}_{\bf j} \nonumber \\
 &+& \! \lambda \sum_{\bf i}
 (Q_{2{\bf i}} \tau_{x{\bf i}} \!+\! Q_{3{\bf i}}\tau_{z{\bf i}})
 \!+\! (1/2) \sum_{\bf i}(Q_{2{\bf i}}^2 \!+\! Q_{3{\bf i}}^2),
\end{eqnarray}
with 
$\tau_{x{\bf i}}$=$\sum_{\sigma}(c_{{\bf i} a\sigma}^{\dag}
c_{{\bf i}b\sigma}$$+$$c_{{\bf i} b\sigma}^{\dag}c_{{\bf i}a\sigma})$
and
$\tau_{z{\bf i}}$=$\sum_{\sigma}
(c_{{\bf i} a\sigma}^{\dag}c_{{\bf i}a\sigma}$
$-$$c_{{\bf i} b\sigma}^{\dag}c_{{\bf i}b\sigma})$,
where $c_{{\bf i}a \sigma}$ ($c_{{\bf i} b \sigma}$) is
the annihilation operator for an $e_{\rm g}$ electron with spin $\sigma$ 
in the $d_{x^2-y^2}$ ($d_{3z^2-r^2}$) orbital at site ${\bf i}$,
${\bf a}$ is the vector connecting nearest-neighbor sites, and 
$t^{\bf a}_{\gamma \gamma'}$ is the hopping amplitude between 
$\gamma$- and $\gamma'$-orbitals in neighboring sites
along the ${\bf a}$-direction, given by 
$t^{\bf x}_{aa}$=$-\sqrt{3}t^{\bf x}_{ab}$=$-\sqrt{3}t^{\bf x}_{ba}$=
$3t^{\bf x}_{bb}$=$3t/4$ for ${\bf a}$=${\bf x}$,
$t^{\bf y}_{aa}$=$\sqrt{3}t^{\bf y}_{ab}$=$\sqrt{3}t^{\bf y}_{ba}$=
$3t^{\bf y}_{bb}$=$3t/4$ for ${\bf a}$=${\bf y}$,
and $t^{\bf z}_{bb}$=$3t/4$,
$t^{\bf z}_{aa}$=$t^{\bf z}_{ab}$=$t^{\bf z}_{ba}$=0 for 
${\bf a}$=${\bf z}$.
The Hund coupling $J_{\rm H}$$(>0)$ links the $e_{\rm g}$ electron spin 
${\bf s}_{\bf i}$=$\sum_{\gamma\alpha\beta}$$c^{\dag}_{{\bf i}\gamma\alpha}$
$\bbox{\sigma}_{\alpha\beta}$$c_{{\bf i}\gamma\beta}$ and the
localized $t_{\rm 2g}$ spin ${\bf S}_{\bf i}$ assumed classical
($|{\bf S}_{\bf i}|$=1). 
$J'$ is the AFM-coupling between nearest-neighbor $t_{\rm 2g}$ spins.
The dimensionless electron-phonon coupling constant is $\lambda$.
$Q_{2{\bf i}}$ and $Q_{3{\bf i}}$ are, respectively, 
the $(x^2$$-$$y^2)$- and $(3z^2$$-$$r^2)$-type JT modes of the MnO$_6$
octahedron.

To simplify the model the widely used limit $J_{\rm H}$=$\infty$ is
considered in this work.
In such a limit, the $e_{\rm g}$ electron spin perfectly aligns along 
the $t_{\rm 2g}$ spin direction, reducing the number of degrees of
freedom.
$\lambda$ is evaluated from $\lambda$=$\sqrt{2E_{\rm JT}/t}$,
where the static JT energy
$E_{\rm JT}$ is estimated as $E_{\rm JT}$=0.25eV\cite{Shen} and $t$ is
typically 0.2eV in narrow-band manganese oxides such as PCMO.
Thus, in this paper $\lambda$ is fixed as $1.6$ to compare the
theoretical predictions with experiments, but the results shown below
are not much sensitive to small changes in $\lambda$.\cite{comment1}
In the rest of the paper, the importance of the $J'$-dependence of
ground state energies is emphasized to address the FM-CO competition.

To solve the Hamiltonian (1), in this paper both a numerical
relaxation technique and an analytic mean-field (MF) approximation are
employed. 
In the former, the optimized JT distortion and $t_{\rm 2g}$ spin
configuration are determined numerically by the Simplex Method.
Although the result is very accurate, considerable CPU time is needed
to achieve convergence, and it is difficult to treat large clusters
with this method. 
In fact, the cluster studied here is a 4$\times$4$\times$4 cube with periodic
boundary conditions (PBC).
However, this lattice size is enough for the present investigation 
at $x$=1/4, 3/8, and 1/2.
If other dopings such as $x$=0.3 and 0.4 are studied, larger size
lattices should be treated due to the complexity of the CO states 
described here, which typically have a large unit cell.
In the analytic approach, on the other hand, the JT distortion is
determined self-consistently at each site using the relations
$Q_{2{\bf i}}$=$-\lambda$$\langle \tau_{x{\bf i}} \rangle$
and 
$Q_{3{\bf i}}$=$-\lambda$$\langle \tau_{z{\bf i}} \rangle$, 
where the bracket denotes the average value.
In this approach, energies for a variety of possible $t_{\rm 2g}$ spin 
patterns are compared to find the lowest-energy state.
This method has the advantage that a large-size lattice can
be treated without much CPU time, but it must be checked whether the
obtained structure indeed corresponds to the lowest-energy state.
Such a check is here carried out by comparing MF and unbiased
numerical results.
Thus, the combination of the analytic and numerical techniques is
powerful to obtain accurate predictions rapidly. 
In addition, the reliability of the relaxation technique has been
checked by comparing data with unbiased 
Monte Carlo (MC) simulations.\cite{Yunoki}
At $x$=0, the agreement between the relaxation technique and MC
simulation was excellent in any dimensions within
errorbars,\cite{Hotta1} indicating the reliability of the present combined
analytic-numeric approach.

\begin{figure}[h]
\vskip1.4truein
\hskip-0.2truein
\centerline{\epsfxsize=3.truein \epsfbox{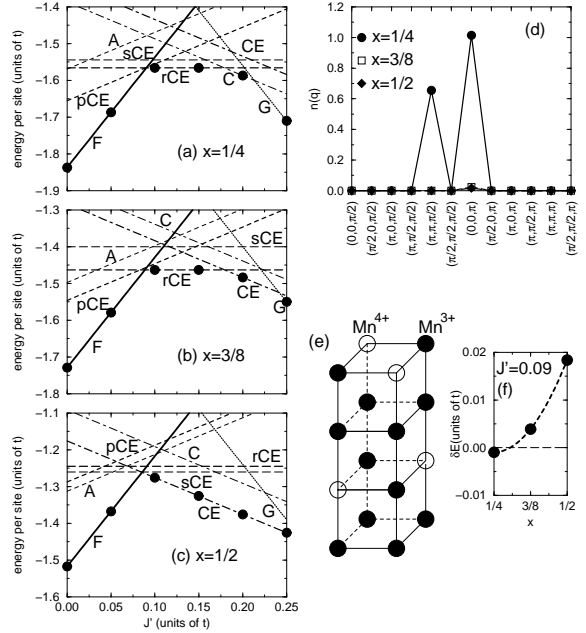} }
\vskip-0.6truein
\caption{
Energies per site vs. $J'$ of several spin arrangements for (a)
$x$=1/4, (b) 3/8, and (c) 1/2 in a 4$\times$4$\times$4 cubic lattice
with PBC.
The solid circles and lines are obtained by the relaxation method and
the MF approximation, respectively.
The meanings of the lines are as follows:
Thick solid (FM), thick broken (pCE), thin broken (A-AFM),
thick dashed (rCE), thin dashed (sCE), thick dot-dashed (CE),
thin dot-dashed (C-AFM), and dotted (G-AFM).
(d) Charge correlation function in the FM phase for $x$=1/4 (solid
circle), 3/8 (open square), and 1/2 (solid diamond).
(e) Schematic charge configuration for the FM insulating phase at
$x$=1/4. To save space, only a 2$\times$2$\times$4 cluster is shown,
but the structure is periodically repeated in all directions. 
(f) Energy difference $\delta E$ between the FM and CO phases vs. $x$
for $J'$=0.09. The broken curve is a spline interpolation.
Note that the region with positive $\delta E$ corresponds to a CO
ground state.
}
\label{fig1}
\end{figure}

Now let us analyze our results for $x$=1/4, 3/8, and 1/2
(Fig.1(a)-(c)), studying energies vs. $J'$ for several spin patterns.
The MF results agree accurately with those of the
relaxation technique, confirming the reliability of the MF method. 
For $J'$$\alt$ 0.1, the FM phase is stabilized and 
it is insulating at $x$=1/4, as shown in Fig.~1(d) where sharp peaks
can be observed in the Fourier transform of the charge correlation
function defined as 
$n({\bf q})$=$(1/N)$$\sum_{\bf i,j}$$e^{-i{\bf q}\cdot({\bf i}-
{\bf j})}$$\langle$$(n_{\bf i}$$-$$n)$$(n_{\bf j}$$-n)$$\rangle$.
Here $N$ is the total number of sites, 
$n_{\bf i}$=$\sum_{\sigma\gamma}$
$c_{{\bf i} \gamma\sigma}^{\dag}c_{{\bf i}\gamma\sigma}$,
and $n$(=1$-$$x$) is the average electron number per site.
The peaks in Fig.~1(d) indicate a CO pattern, schematically 
shown in Fig.~1(e), in which two-dimensional (2D) planes with $n$=1
and 1/2, respectively, are stacked along the $z$-axis.\cite{comment2} 
This charge-ordered FM insulating phase emerging from our calculations 
may be experimentally detected for PCMO.
Note that the ``FM insulating'' phase of manganites has not been
analyzed in detail in previous studies, and here it is conjectured that it
has charge-ordering.
For $x$=3/8 and 1/2, the FM phases are found to be metallic,
since no clear peak can be observed in $n({\bf q})$.
This insulator-metal change in the FM state as a function of $x$ 
agrees quite well with PCMO experiments.

Around $J'$$\approx$0.1, a close competition occurs 
among the FM phases and several CE-type CO states (to be described later), 
indicating that a small perturbation can easily induce
a first-order transition between FM and CO phases.
It should be emphasized the remarkable result 
that such a competing region does $not$
sensitively depend on $x$, suggesting that a CMR phenomenon can occur 
in a wide range of $x$ as observed in PCMO for $x=0.3$$\sim$$0.5$.
In fact, as shown in Fig.~1(f), the energy difference 
between the FM and CO phases are 0.004 at $x$=3/8 and 
0.018 at $x$=1/2 for $J'$=0.09.
If $t$ is assumed to be 0.2eV, those are about 8 and 36 Tesla,
similar values as observed in the experiments.\cite{footnote1}

Let us discuss the detail structure of the CO phase around
$J'$$\sim$0.1.
Our results suggest that the possible CO phases are given by several 
combinations of the 2D CE-type AFM spin pattern.
To simplify the discussion, the CE-type AFM configuration in Fig.~2(a)
is used as the basic pattern.
The possible ground states are classified into four types by the stacking
manner along the $z$-axis.
{\bf (i)}``the planar CE-type'' (pCE), in which the pattern (a)
stacks along the $z$-axis without any change.
{\bf (ii)}``CE-type'' (CE), in which the patterns (a) and (b) stack
along the $z$-axis alternatively. Note that this is the CE-type AFM
structure observed in several half-doped manganites. 
{\bf (iii)}``shifted CE-type'' (sCE), in which patterns (a) and (c) 
stack along the $z$-axis alternatively. Note that (c) is obtained by
shifting (a) by one lattice spacing along the $y$-axis.
This structure, with no charge-stacking, was suggested as a possible
$x$=1/2 ground state, if the NaCl-type CO occurs due to
the strong nearest-neighbor repulsion.\cite{Yunoki}
{\bf (iv)}``rotated CE-type'' (rCE), in which patterns (a) and (d)
stack along the $z$-axis alternatively. Note that (d) is obtained by 
rotating (a) by $\pi$ around a certain corner site in the zigzag FM
chain in the $x$-$y$ plane, or by a two lattice spacing shift. 
Both the rCE and sCE states have the same number of AF and FM links
and their energies are $J'$-independent.

As observed in Figs.1(a)-(c), both the pCE- and rCE-phases compete with
the FM phase for $x$=1/4 and 3/8, while at $x$=1/2, both sCE- and,
especially, the CE-phase are competitive with the FM phase.
As for the CO pattern, the CE-, pCE-, and rCE-phases exhibit
the charge stacking due to a peak 
in $n($${\bf q}$=$(\pi,\pi,0)$$)$ (Fig.~2(e)).
On the other hand, the sCE-phase has the NaCl-type charge ordering
with a peak at ${\bf q}$=$(\pi,\pi,\pi)$.
Since experimentally $z$-axis charge stacking has been observed,
the pCE-, rCE-, and CE-type phases are the best candidates 
for the ground state of the CO phases, since their energies are low
in the analysis reported here.

\begin{figure}[h]
\hskip-2.2truein
\centerline{\epsfxsize=2.truein \epsfbox{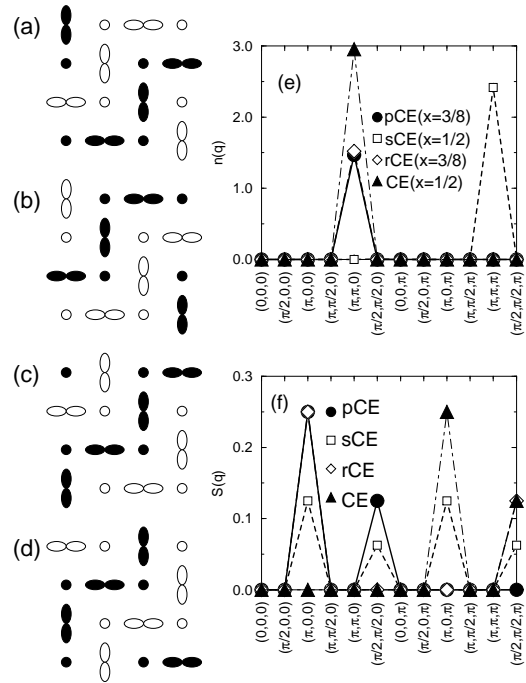} }
\vskip-3.5truein
\caption{(a)-(d) Four types of AFM CE-type spin configurations 
in the $x$-$y$ plane. Solid and open symbols indicate up and down 
$t_{\rm 2g}$ spins, respectively.
The lobes indicates 
$(3x^2-r^2)$- or $(3y^2-r^2)$-orbital at Mn$^{3+}$ site, while
the circles denotes either Mn$^{4+}$ or an imperfectly doped site.
The stacking along the $z$-axis of (a)-(a),
(a)-(b), (a)-(c), and (a)-(d) lead to pCE-, CE-, sCE-, and rCE-phases, 
respectively.
(e) Charge correlation for pCE-, CE-, sCE-, and rCE-phases.
For pCE- and rCE-types, results at $x$=3/8 are shown, while 
for sCE- and CE-types, $x$ is chosen as 1/2.
(f) $t_{\rm 2g}$ spin correlations for pCE-, CE-, sCE-, and rCE-phases.}
\label{fig2}
\end{figure}

Now consider the spin correlations 
$S({\bf q})$=$(1/N)\sum_{\bf i,j}$
$e^{-i{\bf q}\cdot({\bf i}-{\bf j})}$
$\langle$${\bf S}_{\bf i}$$\cdot$${\bf S}_{\bf j} \rangle$.
The spin patterns of the states found by combining Fig.~2(a)-(d) have
the characteristic zigzag FM chains in the $x$-$y$ plane, leading to 
peaks at $(\pi,0,q_z)$ and $(\pi/2,\pi/2,q_z)$, where $q_z$ depends on
the stacking arrangement.
For pCE- and CE-types, $q_z$ is given by 0 and $\pi$, respectively,
due to the FM- and AFM-spin stacking along the $z$-axis (Fig.~2(f)).
For rCE-type, the peaks appear at $(\pi,0,0)$ and $(\pi/2,\pi/2,\pi)$, 
since half the $z$-axis bonds are FM and half AF.
For sCE-type, four peaks exist 
at $(\pi,0,q_z)$ and $(\pi/2,\pi/2,q_z)$ with $q_z$=0 and $\pi$.
In neutron experiments for 
PCMO with $x=0.3$ and 0.4,\cite{Yoshizawa}
a peak was found at ${\bf q}$=$(\pi/2,\pi/2,0)$,
suggesting that the pCE-state is the best candidate for the 
ground state in 0.3 $\alt$$x$$\alt$ 0.4.
The well-defined charge and spin arrangement discussed here
considerably improves over previous more vague
``$x$=1/2 CE-type plus defects'' descriptions of this state.
At $x$=1/2, the $(\pi/2,\pi/2,\pi)$ 
peak\cite{Kajimoto} indicates the CE-type as the ground state.

Note that the present results do not always provide the 
pCE-type as the lowest-energy state in a sizable region 
of $J'$ at $x$=1/4 and 3/8.
To stabilize the pCE-phase it may be necessary to
include the Coulomb interactions, especially the
nearest-neighbor repulsion $V$.
In fact, if the nearest-neighbor charge correlation
$C_{\rm NN}$=
$\sum_{\langle {\bf i,j} \rangle} \langle \rho_{\bf i}\rho_{\bf j} \rangle$
is evaluated, $C_{\rm NN}$ for the pCE phase is found to be smaller
than that of the rCE-type, indicating that rCE-type is indeed more
energetically penalized by $V$. 
Thus, it is reasonable to expect that a stability
``window'' for the pCE-state 
will appear if it were possible to include accurately
the Coulomb interactions.
Note, however, that the size of such a ``window'' will sensitively 
depend on the parameter choice, cluster size, and approximations,
since a small perturbation can easily modify ground state properties
in a region where several states are competing.\cite{comment3}
Moreover, at $x$=1/2, the sCE-phase with the NaCl-type CO will be
stabilized if $V$ is very large, making the situation more
complicated since it is known that charge stacking, not present in the
sCE-phase, occurs at $x$=1/2. Nevertheless,
from our present analysis it can be safely concluded that either the
pCE- or rCE-states can be the lowest-energy state and play the role of
the experimentally observed CO phase in the interesting region of PCMO. 

As for the orbital ordering, the alternation of 
$(3x^2$$-$$r^2)$- and $(3y^2$$-$$r^2)$-orbitals appears in the $x$-$y$
plane, although the exact orbital shape is slightly dependent on the 
stacking manner.
At the corner sites in the FM zigzag path, except for sCE-phase, 
our calculation suggests 
that the orbital is polarized along the $z$-axis and the 
$(3z^2$$-$$r^2)$-orbital is occupied.
This result agrees well with the experimental data.\cite{Jirak,Okimoto}
To distinguish between pCE- and rCE-phases, the stacking of orbitals
along the $z$-axis is important.
Namely, for the pCE and rCE phases, the $(3x^2$$-$$r^2)$- and 
$(3y^2$$-$$r^2)$-orbitals 
stacks in ferro and antiferro patterns, respectively.
This difference can be detected for PCMO at $x$$\sim$3/8
by using the resonant X-ray scattering technique.

Finally, the possibility of spin canting is here briefly  discussed.
For pCE-type at $x$=3/8, the $t_{2g}$ spins canting
between two adjacent $x$-$y$ planes along the $z$-axis was studied
as a possible way to convert continuously from pCE-type to CE-type.
However, when the ground state energy is plotted against
the canting angle, local minima were found only at angles 
corresponding to the extreme cases pCE- and CE-types,
and the spin canting phase was not stabilized in the present work.
However, the existence of the spin canting state cannot be totally
excluded if a finite value of $J_{\rm H}$ 
is considered (here $J_{\rm H}$=$\infty$ was used).

Summarizing, spin-charge-orbital ordering in manganites has been 
investigated at $x$=1/4, 3/8, and 1/2 using analytical 
and numerical techniques.
For fixed values of $\lambda$ and $J'$, 
it has been found that the energy difference between the FM and CO
phases is remarkably small for $x$=1/4, 3/8, and 1/2, and a small
magnetic field can induce a CO-FM transition. 
This result is a first step toward a possible explanation of the 
CMR effect in PCMO which occurs in a robust density range.
Contrary to previous descriptions of these states as ``$x$=1/2 plus
defects'', here it has been shown that specific charge arrangements without
defects can be constructed to represent the CO states at
$x$$\neq$0. Our predictions can be verified using X ray scattering
experiments.

The authors thank H. Yoshizawa for useful conversations.
T.H. is supported by the Ministry of Education, Science,
Sports, and Culture of Japan. 
E.D. is supported by grant NSF-DMR-9814350.


\end{multicols}

\begin{references}
\bibitem{comment1} If $\lambda$ becomes too small or too large,
the results will change qualitatively.
For $\lambda \agt 2$, the FM phase becomes always insulating even at
$x$=1/2, contrary to the metallic properties of this state in the CMR
phenomenon. 
When $\lambda$ is smaller than unity, the FM insulating phase does not
appear even at $x$=1/4, also contrary to experiments for the (Pr,Ca)
based manganite.
Thus, a moderate value of $\lambda$ between 1 and 2 appears reasonable 
for PCMO.
\bibitem{footnote1} For $J'$ between 0.15 and 0.20, the C- and 
CE-type AFM states are obtained for $x$=1/4 and 3/8, respectively, 
but those results are spurious due to the lattice size.
This point will be discussed elsewhere.
\bibitem{comment3} In fact, the CE-type AFM state is $not$ the true
  long-range order at $x$=1/2, as suggested in 
Ref.~\onlinecite{Kajimoto}.
\end{references}
\end{document}